\documentclass[prb,twocolumn]{revtex4}
\usepackage{graphicx}

\begin{document}

\title{Gradient and Amplitude Scattering in Surface-Corrugated Waveguides}

\author{F.~M.~Izrailev}
\email{izrailev@venus.ifuap.buap.mx}

\affiliation{Instituto de F\'{\i}sica,}

\author{N.~M.~Makarov}
\email{makarov@siu.buap.mx}

\affiliation{Instituto de Ciencias,}

\author{M.~Rend\'{o}n}
\email{mrendon@venus.ifuap.buap.mx}

\affiliation{Facultad de Ciencias de la Electr\'onica,\\
Universidad Aut\'{o}noma de Puebla, Puebla, Pue., 72570,
M\'{e}xico}

\date{\today}

\begin{abstract}
We investigate the interplay between amplitude and square-gradient
scattering from the rough surfaces in multi-mode waveguides
(conducting quantum wires). The main result is that for any (even
small in height) roughness the square-gradient terms in the
expression for the wave scattering length (electron mean free
path) are dominant, provided the correlation length of the surface
disorder is small enough. This important effect is missed in
existing studies of the surface scattering.
\end{abstract}


\maketitle

{\bf 1.} The problem of wave transport (both classical and
quantum) through the guiding surface-disordered systems has a
quite long history and remains a hot topic till now (see, e.g.,
Refs.~\onlinecite{BFb79,RytKrTat89,McGM84,SMFY9899,Mig77,Konrady74,%
TJM86,TrAsh88,MeyStep949597989900,LunKyReiKr96,IsPuzFuks9091,%
MT9801,LunMenIz00,VoronB94,KMYa91,MakMorYam95,Tatar93} and
references therein). One of the main tools to treat this problem
is the reduction of the wave/electron {\it surface} scattering to
the {\it bulk} one in such a way that the latter can be described
by an effective Hamiltonian with a complicated potential, however,
with flat boundaries. Applying this approach, one can reasonably
discriminate between the so-called {\it amplitude} and {\it
gradient} scattering, and analyze their interplay explicitly.

To the best of our knowledge, originally the idea of this approach
was discussed by Migdal~\cite{Mig77}. After, it was frequently
used in the theories of classical and quantum wave/electron
scattering,
see, e.g., Refs.~\onlinecite{Konrady74,TJM86,TrAsh88,%
MeyStep949597989900,LunKyReiKr96,IsPuzFuks9091,MT9801,LunMenIz00}.
But in the majority of
them~\cite{Konrady74,TJM86,TrAsh88,MeyStep949597989900,LunKyReiKr96}
the study was restricted to the lowest order in the roughness
height $\sigma$. Other methods
\cite{IsPuzFuks9091,MT9801,LunMenIz00,VoronB94,KMYa91,MakMorYam95,Tatar93}
were mainly based on the principal assumption that the surface
roughness is sufficiently smooth.

In this contribution we present the theory of wave scattering from
rough surfaces, which takes into account both the amplitude and
gradient scattering mechanisms. The important point is that we do
not assume any special restrictions on the model parameters except
for general conditions of {\it weak scattering}. The latter
provides us with an appropriate {\it perturbative} approach in
scattering potential, however, is not restricted by the smoothness
of surfaces.

The main attention is paid to the competition between the
amplitude and gradient scattering. One of our main results is that
at fixed r.m.s. roughness height $\sigma$, the less correlation
length $R_c$ of a random surface profile, the larger contribution
of the gradient mechanism. Thus, by passing from the smooth to
white-noise profiles, the gradient scattering becomes to prevail.
We have analyzed this crossover and obtained the estimates
describing the transition to the dominating gradient scattering.
In particular, we show that this transition is located within the
region of small roughness slopes where $\sigma/R_c\ll1$.

{\bf 2}. In what follows we consider an open plane wavegui\-de (or
conducting quasi-one-dimensional quantum wire) of the average
width $d$, stretched along the $x$-axis. For simplicity, one
(lower) surface of the waveguide is assumed to be flat, $z=0$,
while the other (upper) surface has a rough profile
$z=w(x)=d+\sigma\xi(x)$ with $\left< \xi(x) \right>=0$ and $\left<
\xi^2(x) \right>=1$. The average $\left< ... \right>$ is performed
over different realizations of a statistically homogeneous and
isotropic {\it Gaussian} random function $\xi(x)$. We also assume
that its pair correlator $\langle\xi(x)\xi(x')\rangle={\cal
W}(x-x')$ decreases on a scale $R_c$, with the normalization
${\cal W}(0)=1$. The roughness-height power (RHP) spectrum
$W(k_x)$ is defined by
\begin{equation}\label{FT-W}
W(k_x)=\int_{-\infty}^{\infty}dx\exp(-ik_xx)\,{\cal W}(x).
\end{equation}
Since ${\cal W}(x)$ is an even function of $x$, its Fourier
transform (\ref{FT-W}) is even, real and nonnegative function of
$k_x$. The RHP spectrum has maximum at $k_x=0$ with $W(0)\sim
R_c$, and decreases on the scale $R_c^{-1}$.

In order to analyze the surface scattering for our model, we
employ the method of the {\it retarded Green's function} ${\cal
G}(x,x';z,z')$. Specifically, we start with the Dirichlet
boundary-value problem
%
\begin{eqnarray}\label{1RB-GFin}
&&\left(\frac{\partial^2}{\partial x^2}+
\frac{\partial^2}{\partial z^2}+k^2\right){\cal G}(x,x';z,z')
\nonumber\\[6pt]
&&=\delta(x-x')\delta(z-z'),
\label{1RB-Geq1}\\[6pt]
&&{\cal G}(x,x';z=0,z')={\cal G}(x,x';z=w(x),z')=0.\qquad\quad
\label{1RB-BC1}
\end{eqnarray}
Here the wave number $k$ is equal to $\omega/c$ for an
electromagnetic wave of the frequency $\omega$ and TE
polarization, propagating through a waveguide with perfectly
conducting walls. As for an electron quantum wire, $k$ is the
Fermi wave number within the isotropic Fermi-liquid model. In
order to express the surface scattering as a bulk one, we perform
the transformation to new coordinates,
\begin{equation}\label{1RB-CoorTr}
x_{new}=x_{old}, \qquad z_{new}=z_{old}\, d/[d+\sigma\xi(x)],
\end{equation}
in which both waveguide surfaces are flat. Correspondingly, we
introduce the canonically conjugate Green's function, ${\cal
G}_{new}=d^{-1}\sqrt{w(x)w(x')}{\cal G}_{old}$ and omit the
subscript ``new'' in what follows. As a result, we arrive at the
equivalent boundary-value problem governed by the equation with a
``bulk" perturbation potential,
%
\begin{eqnarray}\label{1RB-GFbulk}
&&\Bigg\{\frac{\partial^2}{\partial x^2}+
\frac{\partial^2}{\partial z^2}+k^2
-\bigg[1-\frac{d^2}{w^2(x)}\bigg]\frac{\partial^2}{\partial z^2}
\nonumber\\[6pt]
&&-\frac{\sigma}{w(x)}\bigg[\xi'(x)\frac{\partial}{\partial
x}+\frac{\partial}{\partial x}\,\xi'(x)\bigg]\bigg[\frac{1}{2}+
z\frac{\partial}{\partial z}\bigg]
\nonumber\\[6pt]
&&+\frac{\sigma^2{\xi'}^2(x)}{w^2(x)}
\bigg[\frac{3}{4}+3z\,\frac{\partial}{\partial z}+
z^2\frac{\partial^2}{\partial z^2}\bigg]\Bigg\}\,{\cal
G}(x,x';z,z')
\nonumber\\[6pt]
&&=\delta(x-x')\delta(z-z'),
\label{1RB-Geq4}\\[6pt]
&&{\cal G}(x,x';z=0,z')={\cal G}(x,x';z=d,z')=0. \label{1RB-BC4}
\end{eqnarray}
Here the prime stands for the derivative over $x$.

We emphasize that Eq.~(\ref{1RB-GFbulk}) is {\it exact} and valid
for any form of $w(x)$. As one can see, the scattering potential
depends {\it both} on the roughness profile $\sigma\xi(x)$ and on
its gradient $\sigma\xi'(x)$. Moreover, the potential contains the
term with the {\it square gradient} $\sigma^2{\xi'}^2(x)$. This
term is proportional to $\sigma^2$ and for this reason was
neglected in all previous studies of transport properties in the
surface-disordered waveguides. However, as a matter of fact, the
square gradient introduces the operator $\hat{\cal
V}(x)={\xi'}^2(x)-\langle{\xi'}^2(x)\rangle$, which plays a
special role. Its pair correlator,
\begin{equation}\label{V-corr}
\langle {\hat{\cal V}(x)}{\hat{\cal V}(x')}\rangle=
2\langle\xi'(x)\xi'(x')\rangle^2=2{{\cal W}''}^2(x-x'),
\end{equation}
determines the square-gradient power (SGP) spectrum
\begin{equation}\label{T-def}
T(k_x)=\int_{-\infty}^{\infty}dx\exp{(-ik_xx)}\,{{\cal W}''}^2(x).
\end{equation}
One should stress that although by integration by parts the power
spectrum of the roughness gradients $\sigma\xi'(x)$ can be reduced
to the RHP spectrum $W(k_x)$, this is not possible for the SGP
spectrum $T(k_x)$. This very fact reflects a highly non-trivial
role of the {\it square-gradient scattering}, giving rise to the
competition with the well known amplitude scattering, in spite of
the seeming smallness of the term $\sigma^2{\xi'}^2(x)$.

To proceed, we pass from Eq.~(\ref{1RB-GFbulk}) to the Dyson-type
equation, performing the ensemble averaging with the use of the
technique developed in Ref.~\onlinecite{McGM84}. The method allows
one to develop the consistent perturbative approach with respect
to the scattering potential, which takes adequately into account
the {\it multiple scattering} from the corrugated boundary. After
quite cumbersome calculations we have obtained the average Green's
function which in the normal-mode representation has the form
\begin{eqnarray}
&&\langle{\cal G}(x,x';z,z')\rangle=\sum_{n=1}^{N_d}
\sin\left(\frac{\pi nz}{d}\right) \sin\left(\frac{\pi
nz'}{d}\right)
\nonumber\cr\\
&&\times\frac{\exp(ik_n|x-x'|)}{ik_nd}\,
\exp\left(-\frac{|x-x'|}{2L_n}\right).
\label{avGF}
\end{eqnarray}
Here $k_n=\sqrt{k^2-(\pi n/d)^2}$ corresponds to the unperturbed
lengthwise wave number $k_x$, and $N_d=[kd/\pi]$ is the number of
propagating modes (or conducting electron channels) determined by
the integer part $[\ldots]$ of the ratio $kd/\pi$.

{\bf 3.} Our interest is in the {\it attenuation length} or {\it
total mean free path} $L_n$ of the $n$-th mode. Its inverse value
is given by the imaginary part of the proper self-energy and, in
accordance with the form of the scattering potential, consists of
two terms describing different scattering mechanisms,
\begin{equation}\label{1RB-Ln-sum}
\frac{1}{L_n}=\frac{1}{L^{(1)}_n}+\frac{1}{L^{(2)}_n}.
\end{equation}
The first length $L^{(1)}_n$ is determined by the expression
\begin{eqnarray}
\frac{1}{L^{(1)}_n}&=&\sigma^2\frac{(\pi
n/d)^2}{k_nd}\,\sum_{n'=1}^{N_d}\frac{(\pi n'/d)^2}{k_{n'}d}
\nonumber\\
&&\times\left[ W(k_n+k_{n'})+W(k_n-k_{n'})\right].
\qquad\qquad\label{1RB-Ln1-def}
\end{eqnarray}
Its {\it diagonal term} is formed by the {\it amplitude}
scattering while the {\it off-diagonal terms} result from the {\it
gradient} one. Eq.~(\ref{1RB-Ln1-def}) coincides with that
previously obtained by different methods (see, e.g.,
Ref.~\onlinecite{BFb79}).

The second length $L^{(2)}_n$ is associated solely with the {\it
square-gradient} mechanism due to the operator $\hat{\cal V}(x)$,
\begin{equation}\label{1RB-LnG-mode}
\frac{1}{L^{(2)}_n}=\sum_{n'=1}^{N_d}\frac{1}{L^{(2)}_{nn'}}.
\end{equation}
Its diagonal term controls the {\it intramode scattering},
\begin{eqnarray}\label{1RB-LnnG}
\frac{1}{L^{(2)}_{nn}}&=&\frac{\sigma^4}{2}\frac{(\pi
n/d)^4}{k_n^2}\,\left[\frac{1}{3}+\frac{1}{(2\pi n)^2}\right]^2
\nonumber\\
&&\times\left[T(2k_n)+T(0)\right].
\end{eqnarray}
The off-diagonal partial length $L^{(2)}_{n\neq n'}$ describes the
{\it intermode scattering} (from $n$ to $n'\neq n$ channels),
\begin{eqnarray}\label{1RB-Lnn'2}
\frac{1}{L^{(2)}_{n\neq n'}}&=&\frac{8\sigma^4}{\pi^4}\frac{(\pi
n/d)^2}{k_n}\,\frac{(\pi n'/d)^2}{k_{n'}}\,
\frac{(n^2+n'^2)^2}{(n^2-n'^2)^4}
\nonumber\\
&&\times\left[T(k_n+k_{n'})+T(k_n-k_{n'})\right].
\end{eqnarray}
To the best of our knowledge, in the surface-scattering problem
for multi-mode waveguides the operator ${\hat{\cal V}(x)}$ was
never taken into account, and, as a result, the square-gradient
attenuation length $L^{(2)}_n$ was missed in previous studies.

Let us analyze the conditions under which Eqs.~(\ref{1RB-Ln-sum})
-- (\ref{1RB-Lnn'2}) are derived. We stress that the Dyson-type
equation for the average Green's function was obtained within the
second-order approximation in the perturbation potential. This
means that the self-energy in this equation contains the binary
correlator of the surface-scattering potential and the unperturbed
Green's function. In terms of the diagrammatic technique this is
similar to the ``simple vortex" or, the same, Bourret
approximation~\cite{Bourret62}. Following the ideas discussed in
Ref.~\onlinecite{RytKrTat89}, one can show that this approximation
is justified when the broadening $1/2L_n$ of the quantum wave
number $k_n$ is much less than both the correlation scale
$R_c^{-1}$ and the spacing $|k_n-k_{n\pm1}|\approx|\partial
k_n/\partial n|$ between neighboring quantum wave numbers. The
same conditions also arise due to another approximation which is
the use of the unperturbed value $k_n$ in the expression for the
self-energy, instead of the perturbed one. Now we take into
account that $|\partial k_n/\partial n|\sim\Lambda_n^{-1}$, where
$\Lambda_n$ is the distance between two successive reflections of
a wave from the rough boundary inside the $n$-th channel. As a
result, we come to the following conditions of a {\it weak surface
scattering}
\begin{equation}\label{WS-Ln}
\Lambda_n=2k_nd/(\pi n/d) \ll2L_n, \qquad R_c\ll2L_n.
\end{equation}
These inequalities imply that the electron/wave weakly attenuates
on both the correlation length $R_c$ and the {\it cycle length}
$\Lambda_n$.

As one can see, the expressions (\ref{1RB-Ln1-def}) and
(\ref{1RB-LnG-mode}) -- (\ref{1RB-Lnn'2}) represent, respectively,
basic contributions from principally different surface-scattering
mechanisms related to the {\it amplitude, gradient} and {\it
square-gradient} terms. It should be emphasized that the
corrections proportional to $\sigma^4$, originated from higher
order approximations in the {\it amplitude} and {\it gradient}
terms of the perturbation potential, are smaller than the main
contribution (\ref{1RB-Ln1-def}) under the conditions
(\ref{WS-Ln}). Contrary, the {\it square-gradient} terms give rise
to the $\sigma^4$-terms in Eqs.(\ref{1RB-LnnG})-(\ref{1RB-Lnn'2})
which should not be neglected due to a specific dependence on the
correlation length $R_c$. Note that Eq.~(\ref{WS-Ln}) implicitly
includes the requirement for the surface corrugations be small in
height ($\sigma\ll d$), but does not restrict the value
$\sigma/R_c$ of the roughness slope.

{\bf 4.} Since $L^{(1)}_n$ and $L^{(2)}_n$ depend on as many as
four dimensionless parameters $(k\sigma)^2$, $kR_c$, $kd/\pi$, and
$n$, the complete analysis appears to be quite complicated. For
this reason, below we restrict ourselves by the analysis of the
interplay between $L^{(1)}_n$ and $L^{(2)}_n$ as a function of the
dimensionless correlation length $kR_c$ for $N_d\approx kd/\pi\gg
1$.

As follows from Eq.~(\ref{1RB-Ln1-def}), the inverse value of the
amplitude-scattering length typically increases with an increase
of $kR_c$. Specifically, in the case of the small-scale roughness
($kR_c\ll 1\lesssim k\Lambda_n$) we have $1/L^{(1)}_n\propto
kR_c$. Then, within the intermediate region where $1\ll kR_c\ll
k\Lambda_n$, the increase of $1/L^{(1)}_n$ slows down, or can even
be replaced by the decrease for some values of the model
parameters. Finally, for large-scale roughness and strong
correlations ($1\lesssim k\Lambda_n\ll kR_c$) the value of
$1/L^{(1)}_n$ again starts to increase linearly with $kR_c$.

In contrast with $1/L^{(1)}_n$, the inverse square-gradient
scattering length $1/L^{(2)}_n$ reveals a monotonous decrease as
the parameter $kR_c$ increases. At small ($kR_c\ll 1\lesssim
k\Lambda_n$) and extremely large ($1\lesssim k\Lambda_n\ll kR_c$)
values of $kR_c$, this decrease obeys the law
$1/L^{(2)}_n\propto(kR_c)^{-3}$, due to $T(0)\sim R_c^{-3}$.

>From this analysis it becomes clear that the curves displaying
$1/L^{(1)}_n$ and $1/L^{(2)}_n$ must intersect, and one can
observe the crossover from the square-gradient to amplitude
surface scattering. To the left from the crossing point
$(kR_c)_{\odot}$ the square-gradient scattering length prevails,
$L^{(2)}_n\ll L^{(1)}_n$. To its right the main contribution is
due to the well known amplitude scattering, $L^{(1)}_n\ll
L^{(2)}_n$. If the crossing point falls onto the interval of the
small-scale roughness ($kR_c\ll1$), its dependence on the model
parameters is described by
\begin{equation}\label{cp-SSR}
(kR_c)_{\odot}^2\sim (k\sigma)n/\sqrt{k_nd}.
\end{equation}
This estimate shows that the crossing point is smaller for smaller
values of the dimensionless roughness height $k\sigma$, as well as
for smaller mode indices $n$, or for larger values of the
parameter $kd/\pi$.

\begin{figure}[htb]
\includegraphics[angle=270,width=\columnwidth]{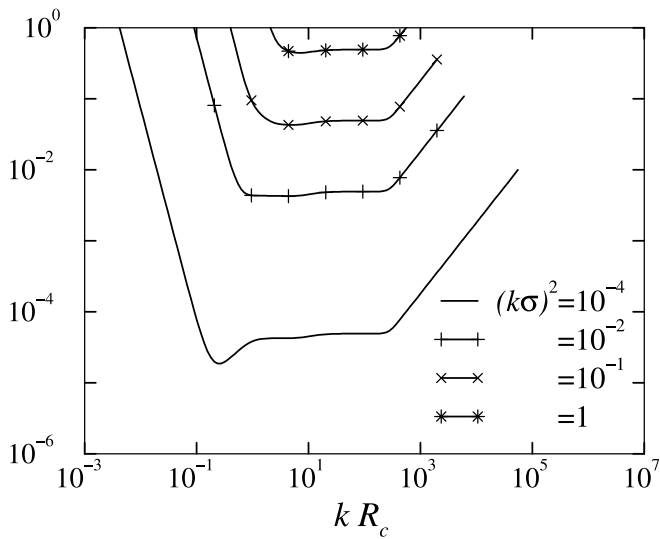}
\caption{\label{fig:1RB_al_L1L2_kRc} $\Lambda_n/2L_n$ versus
$kR_c$ at $k d/\pi=62.5$, $n=31$ and different $(k\sigma)^2$.}
\end{figure}

In Fig.~\ref{fig:1RB_al_L1L2_kRc} we display the dependence of
$\Lambda_n/2L_{n}$ as a function of $kR_c$ assuming the Gaussian
binary correlator ${\cal W}(x)=\exp(-x^2/2R_c^2)$ for random
surface profile $\xi(x)$. The curves are plotted starting from
such values of $kR_c$ for which $\Lambda_n/2L_n^{(2)}=1$,
according to the first condition of Eq.~(\ref{WS-Ln}). Taking into
account the second condition restricting the maximal value of
$kR_c$, we plot every curve within the range where
$R_c<2L^{(1)}_n$. As one can see, all curves have the crossover
from the square-gradient to amplitude surface scattering. The
first (lowest) one with $(k\sigma)^2=10^{-4}$ has the crossing
point $(kR_c)_{\odot}\approx 0.2$ located within the interval of
small-scale roughness, and the crossover reveals a small dip
centered at $(kR_c)_{\odot}$. The curve obeys the asymptotic
behavior $(kR_c)^{-3}$ to the left from $(kR_c)_{\odot}$ due to
the main contribution from $\Lambda_n/2L_n^{(2)}$. Then the
quantity $\Lambda_n/2L_n^{(1)}$ becomes dominating in the sum
(\ref{1RB-Ln-sum}), therefore, the curve starts to rise. Firstly,
the linear dependence on $kR_c$ on the right deep-side (where
$kR_c<1$) is replaced with a smoother one (for $kR_c>1$). Finally,
for $R_c>\Lambda_n$ (strong correlations) the linear dependence
restores.

The crossing points of the second, third and fourth curves have
the values of the order one. Here the total attenuation length
$L_n$ within the whole small-scale region is formed by the
square-gradient scattering length $L_n^{(2)}$. In full agreement
with Eq.~(\ref{cp-SSR}) the presented curves display that the
smaller parameter $(k\sigma)^2$ the smaller value of the crossing
point $(kR_c)_{\odot}$.

Note that for all curves in Fig.1 the roughness {\it height} is
small, $\sigma/d\ll1$. Furthermore, for the amplitude-dominated
scattering (to the right from the point $(kR_c)_{\odot}$ where
$\Lambda_n/2L^{(1)}_{n}$ mainly contributes), the average
corrugation {\it slope} is also small for all data,
$\sigma/R_c\ll1$. The roughness slope remains to be small at the
crossing points too, but increases to their left with the decrease
of $kR_c$. As a result, to the left from the crossing point where
the square-gradient term $\Lambda_n/2L^{(2)}_{n}$ prevails, the
slope reaches the values of the order one, or even larger for the
first tree curves.

In conclusion, we have discovered the principal importance of the
square-gradient surface-scattering mechanism which was never taken
into account in the literature. We have shown that at any fixed
value of the roughness height $\sigma$ one can indicate the region
of small values of the correlation length $R_c$ where the new
square-gradient scattering length $L^{(2)}_{n}$ predominates over
the known amplitude scattering length $L^{(1)}_{n}$
($L^{(2)}_{n}\ll L^{(1)}_{n}$). The predominance occurs in spite
of the fact that $1/L^{(1)}_{n}$ is proportional to $\sigma^2$
while $1/L^{(2)}_{n}$ is proportional to $\sigma^4$. This happens
since the two lengths are determined by the substantially
different roughness-height $W(k_x)$ and roughness-square-gradient
$T(k_x)$ power spectra, that have vastly different dependencies on
$R_c$. It is remarkable that the square-gradient mechanism
prevails in the commonly used region $kR_c\ll 1$ of a {\it
small-scale boundary perturbation}, where the surface roughness is
typically described via the white-noise potential.

This research was partially supported by the CONACYT (M\'exico)
grant No~43730, and by the VIEP-BUAP (M\'exico) under the grant II
104-05/ING/G.



\end{document}